\documentclass[sn-mathphys-num]{sn-jnl}% Math and Physical Sciences Numbered Reference Style 
\usepackage{tikz}
\usepackage{adjustbox}
\usepackage{lipsum}
\usepackage{graphicx}%
\usepackage{subcaption} % Make sure this line is present
\usepackage{multirow}%
\usepackage{amsmath,amssymb,amsfonts}%
\usepackage{amsthm}%
\usepackage{mathrsfs}%
\usepackage[title]{appendix}%
\usepackage{xcolor}%
\usepackage{textcomp}%
\usepackage{manyfoot}%
\usepackage{booktabs}%
\usepackage{algorithm}%
\usepackage{algorithmicx}%
\usepackage{algpseudocode}%
\usepackage{listings}%

\theoremstyle{thmstyleone}%
\theoremstyle{thmstyletwo}%

\theoremstyle{thmstylethree}%

\begin{document}

\title[Article Title]{Preemptive Holistic Collaborative System and Its Application in Road Transportation}


\author[1,2]{\fnm{Yuan} \sur{Li}}\email{liyuan\_mm@chd.edu.cn}

\author[1,2]{\fnm{Tao} \sur{Li}}\email{lt1186166770@163.com}

\author[1,2]{\fnm{Xiaoxue} \sur{Xu}}\email{2112534668@qq.com}

\author[1,2]{\fnm{Xiang} \sur{Dong}}\email{1290740787@qq.com}

\author[1,2]{\fnm{Yincai}  \sur{Cai}}\email{1304584578@qq.com}

\author*[1,2]{\fnm{Ting} \sur{Peng}}\email{t.peng@ieee.org}

\affil*[1]{\orgdiv{Key Laboratory for Special Area Highway Engineering of Ministry of Education}, \orgname{Chang'an University}, \orgaddress{\street{Middle Section of NanErHuan Road}, \city{Xi'an}, \postcode{710064}, \state{Shaanxi}, \country{China}}}

\affil[2]{\orgdiv{Highway School}, \orgname{Chang'an University}, \orgaddress{\street{Middle Section of NanErHuan Road}, \city{Xi'an}, \postcode{71064}, \state{Shaanxi}, \country{China}}}

%\affil[3]{\orgdiv{Department}, \orgname{Organization}, \orgaddress{\street{Street}, \city{City}, \postcode{610101}, \state{State}, \country{Country}}}


\abstract{Numerous real-world systems, including
manufacturing processes, supply chains, and robotic systems,
involve multiple independent entities with diverse objectives. The
potential for conflicts arises from the inability of these entities to
accurately predict and anticipate each other's actions. To address
this challenge, we propose the Preemptive Holistic Collaborative
System (PHCS) framework. By enabling information sharing and
collaborative planning among independent entities, the PHCS
facilitates the preemptive resolution of potential conflicts. We
apply the PHCS framework to the specific context of road
transportation, resulting in the Preemptive Holistic Collaborative
Road Transportation System (PHCRTS). This system leverages
shared driving intentions and pre-planned trajectories to optimize
traffic flow and enhance safety. Simulation experiments in a two-lane
merging scenario demonstrate the effectiveness of PHCRTS,
reducing vehicle time delays by 90\%, increasing traffic capacity by
300\%, and eliminating accidents. The PHCS framework offers a
promising approach to optimize the performance and safety of
complex systems with multiple independent entities.}




\maketitle

\section{Introduction}\label{sec1}

In the spectrum of systems that span across production and
daily life, there exists a multitude of independent entities, each
with unique attributes and specific action objectives. These
entities, while striving to fulfill their individual goals, often
encounter potential conflicts due to the unpredictability of other
entities' intentions. This unpredictability can result in operational
delays, reduced efficiency, and in extreme cases, lead to severe
consequences\cite{Peng2024}.

Addressing this challenge, this paper introduces the Preemptive Holisic Collaborative System (PHCS), a
novel framework designed to mitigate conflicts through
proactive information sharing and collaborative optimization.
The PHCS integrates an information-sharing mechanism that
encompasses the intentions and action plans of all entities within
a unified system. By leveraging this mechanism, the PHCS is
capable of preemptively orchestrating the future spatio-temporal
trajectories for all entities, thereby averting potential conflicts
and enhancing the system's operational efficiency. This
approach not only reduces the likelihood of conflicts but also has
the potential to eliminate them entirely.

The PHCS has been extended to the domain of road
transportation, culminating in the Preemptive Holistic Collaborative Road Transportation System (PHCRTS). This system
facilitates the exchange of driving intentions among vehicles and
pre-plans their trajectories to ensure seamless and conflict-free
movement. The PHCRTS represents a significant advancement
in traffic management, as it not only optimizes traffic flow and
increases throughput but also enhances safety by virtually
eliminating the risk of traffic accidents. The implementation of
the PHCRTS signifies a paradigm shift in road transportation,
aligning with the growing emphasis on intelligent transportation
systems (ITS) that leverage advanced communication
technologies, data analytics, and predictive modeling to
optimize traffic dynamics and ensure safe, efficient travel.

The PHCRTS is underpinned by cutting-edge research in
ITS, which includes the development of vehicle-to-vehicle
(V2V) and vehicle-to-infrastructure (V2I) communication
protocols, the application of machine learning algorithms for
trajectory prediction, and the integration of real-time traffic data
to inform dynamic route planning. These technologies, when
synergized within the PHCRTS framework, offer a robust
solution to the complex challenges of modern transportation
systems, where the independent actions of numerous entities
must be harmonized for the collective benefit of all road users.

\section{Methodology}
\subsection{System Framework}

The PHCS system is composed of multiple subsystems, each housing a manager and numerous entities. The sub - system manager is responsible for a specific spatial range, and these ranges are seamlessly connected, enabling information sharing among managers. Entities and managers share the same clock, which serves as the basis for scheduling tasks. When an entity has a new intention, it immediately communicates this to the relevant manager based on the manager's spatial jurisdiction.

In this research, PHCS is proposed.
The entire system consists of multiple subsystems.
Each subsystem include a manager and multiple entities,  the manager in subsystems is in charge of a certain spatial range. The managers and the entities have the same clock, in this way, they can executed  their manager approved task according to the timetable with the help of the clock.

These spatial ranges connected seamlessly, information are shared between these managers. The Each entity has its own intention from time to time. When a entity in the subsystem has a new intention, it share the intention to the corresponding manager according its responsible spatial range immediately. The manager solve the conflicts between the intention tasks from all the entities in this spatial range, and try to make all the entities fullfill their tasks efficiently. The planned tasks and the revise tasks are the consense of the subsystem, they are feedback to the corresponding entities once the manager finished the planning work. The entities execute the manager approved tasks step by step, these tasks are detailed actions of the corresponding entity at any given timestamp. In this way, the conflicts between the entities are all elemelated, the efficiency of the entire system and each entity are ensured. If a entity enters the  spatial range of another manager, the manager of the corresponding spatiall range take over the responsibility to approve the intentions of the entity.

In PHCS, the intention of all the entities are shared via the manager. The managers have holistic view of all the entities, they plan all the intention tasks of all the entities, and send the planned tasks to the corresponding entity. If new intentions are sent to the manager, the manager solve all the conflicts between the task and other planned tasks, the tasks in the fronzen cannot be altered, the tasks in critical zone, planning zone can be altered if needed. In this way, all the action of all the entities are approved by the manager beforehand, no conflicts occurs. The manager only alter the future action speed of the entity minorly, the tasks of all the entities can fullfilled in most cases efficiently. All the entities act according to the task schedule, detail spatiotemproal motion information is provided in it, the communiction and collaboration cost is kept low.

In the system, the intentions of entities are decomposed into tasks. To achieve preemptive collaboration, each task is pre-shared with the corresponding manager. Consequently, the sharing time and the execution time of a task differ, and this difference should be sufficiently large to guarantee that the manager has ample time to resolve potential conflicts among these tasks and return the conflict-free planned results to the entities.

\subsection{Spatial and Temporal Coordination Basics}
The PHCS system is composed of multiple subsystems, each housing a manager and numerous entities. The sub - system manager is responsible for a specific spatial range, and these ranges are seamlessly connected, enabling information sharing among managers. Entities and managers share the same clock, which serves as the basis for scheduling tasks. When an entity has a new intention, it immediately communicates this to the relevant manager based on the manager's spatial jurisdiction. 

The intention is then decomposed into tasks, and these tasks need to be shared with the manager well in advance of their execution time. As shown in Fig. \ref{fig:temporal_zone}, the time - task space is divided into five zones: history, frozen, critical, planning, and intention. Only tasks in the intention and planning zones can be submitted to the manager, ensuring that the manager has sufficient time to plan and resolve potential conflicts. This temporal - zone - based mechanism is fundamental for preemptive collaboration in the system.

%Framework of PHCS
\begin{figure}[htbp]
    \centering
    \input{temporal_zone.tex}
    \caption{Temporal Zones of Preemptive Collaboration}
    \label{fig:temporal_zone}
\end{figure}

\subsection{Intention Tasks and Temoral Zones}

If an entity has a intention task wich will be executed at timestamp $t_{intention}$, it should  be share to the manager before timestamp of $t_{intention}-t_{frozen}-t_{critical}$. Otherwise, the manager may have not enough to time to resolve the conflict related to this task. If an entity has an intention task, it should share the task to the manager as soon as possible, in this way, the manager has enough time to resolve possible confilict between the tasks of the entities, additionally, the efficiency  of the individual entities and the whole system can also be improved.

As illustrated in Fig. \ref{fig:temporal_zone}, the horizontal axis represents the system time \(t\), while the vertical axis denotes the execution time of the corresponding task \(\tau\). The entire area is thus partitioned into five zones, namely the history zone, the frozen zone, the critical zone, the planning zone, and the intention zone. The boundary equation between the history zone and the frozen zone is \(\tau = t\); between the critical zone and the frozen zone, it is \(\tau = t + t_{\text{frozen}}\); between the critical zone and the planning zone, \(\tau = t + t_{\text{frozen}} + t_{\text{critical}}\); and between the planning zone and the intention zone, \(\tau = t + t_{\text{frozen}} + t_{\text{critical}} + t_{\text{planning}}\). In Fig. \ref{fig:temporal_zone}, \(t_{\text{frozen}}\) is set to \(10\) seconds, \(t_{\text{critical}}\) is set to \(3\) seconds, and \(t_{\text{planning}}\) is set to \(17\) seconds.

Only the intention zone and the planning zone permit task submissions. That is, only when \(\tau \geq t + t_{\text{frozen}} + t_{\text{critical}}\) is met can the tasks be shared with the system manager. In Fig. \ref{fig:temporal_zone}, the coordinates of task \(T_A\) are \((5, 40)\), indicating that the task is  shared to the system at the \(5\)-second timestamp and its start time is at the \(40\)-second timestamp. It is in the intention zone when it is shared to the manager.  We draw a horizontal line through task $T_A$ in Fig. \ref{fig:temporal_zone}, it crosses with the divider lines, draw vertical lines through the cross points, the lines cross with the horizontal axis. according to the cross points on the horizontal axis, timestamp of start planning, planning start deadline, planning finish deadline, execution for task $T_A$ are found, according to the result, the corresponding timestamps are $10$, $27$, $30$, $40$ second, separately. The coordintates of task $T_B$ is $(20, 40)$, it is shared to the manager at $20$ second, it is in the planning zone. Then manager should start the planning work for task $T_B$ once it is shared. The planning work should start before the planning start deadline, it is $27$ second in this case. Thhe planning work should be finished before the planning finish deadline, $30$ second in this case.

In the frozen zone, all the scheduled actions can not be altered, they are concrete consense among all the entities and the managers. The system just execute all actions and report any abnormal situations occured. If any error occur in this process, the  manager will have to alter the approved tasks and solve this abnormal situation.

\begin{algorithm}
\caption{Sub - System Manager Class}\label{alg:sub_system_manager}
\begin{algorithmic}[1]
\State \textbf{Private Fields:}
\State $\text{configuration} \gets \text{space information, edge of the corresponding spatial domain}$
\State $\text{approved\_task} \gets \text{entity}_1\text{: task}_1\text{, task}_2\text{, ..., task}_n\text{,}$
\State $\hspace{1cm}\text{entity}_2\text{: task}_1\text{, task}_2\text{, ..., task}_n\text{,}$
\State $\hspace{1cm}\text{...}$
\State $\hspace{1cm}\text{entity}_n\text{: task}_1\text{, task}_2\text{, ..., task}_n\text{}$
\State $\text{Connected\_Neighbors} $

\Function{try\_approve}{$\text{Intention<entity, a series of tasks>}$}
    \State $\text{success} \gets \text{False}$
    \While{$\neg\text{success}$}
        \ForAll{$\text{task} \in \text{Intention.tasks}$}
            \If{$\text{task}$ conflicts with $\text{approved\_task}$}
                \State $\text{altered\_intention, influenced} \gets \text{alter(intention)}$
                \State \textbf{break}
            \EndIf
        \EndFor
        \State $\text{success} \gets \text{True}$
    \EndWhile
    \State $\text{approved\_task.add(altered\_intention)}$
    \State $\text{approved\_task.update(influenced)}$
    \ForAll{$\text{entity} \in \text{union(altered\_intention, influenced)}$}
        \State $\text{send(entity, entity.approved\_intention)}$
    \EndFor
\EndFunction

\Function{Run}{}
    \While{$\text{True}$}
        \State $\text{new\_intention} \gets \text{Check\_New\_Intention()}$
        \State $\text{thread} \gets \text{Thread(try\_approve(new\_intention))}$
        \State $\text{thread.start()}$
    \EndWhile
\EndFunction
\end{algorithmic}
\end{algorithm}

\begin{algorithm}
\caption{Alter Function for Sub - System Manager}\label{alg:alter_function}
\begin{algorithmic}[1]

\Function{alter}{$\text{intention}$}
    \State $\text{altered\_intention} \gets \text{copy(intention)}$
    \State $\text{influenced\_entities} \gets \varnothing$
    \ForAll{$\text{task} \in \text{altered\_intention.tasks}$}
        \ForAll{$\text{approved\_entity} \in \text{approved\_task.keys()}$}
            \ForAll{$\text{approved\_task} \in \text{approved\_task[approved\_entity]}$}
                \If{$\text{is\_conflicting(task, approved\_task)}$}
                    \State $\text{new\_task} \gets \text{modify\_task(task, approved\_task)}$
                    \State $\text{altered\_intention.replace(task, new\_task)}$
                    \State $\text{influenced\_entities.add(approved\_entity)}$
                    \State $\text{influenced\_entities.add(altered\_intention.entity)}$
                \EndIf
            \EndFor
        \EndFor
    \EndFor
    \State \Return $\text{altered\_intention, influenced\_entities}$
\EndFunction

\Function{is\_conflicting}{$\text{task1, task2}$}
    \State $\text{conflict\_condition1} \gets \text{task1.start\_time} \in \text{task2.time\_interval}$
    \State $\text{conflict\_condition2} \gets \text{task1.end\_time} \in \text{task2.time\_interval}$
    \State $\text{conflict\_condition3} \gets \text{task2.start\_time} \in \text{task1.time\_interval}$
    \State $\text{conflict\_condition4} \gets \text{task2.end\_time} \in \text{task1.time\_interval}$
    \State $\text{spatial\_conflict} \gets \text{task1.location} = \text{task2.location}$
    \State \Return $(\text{conflict\_condition1} \vee \text{conflict\_condition2} \vee \text{conflict\_condition3} \vee \text{conflict\_condition4}) \wedge \text{spatial\_conflict}$
\EndFunction

\Function{modify\_task}{$\text{task, conflicting\_task}$}
    \State $\text{new\_start\_time} \gets \text{conflicting\_task.end\_time} + 1$
    \State $\text{new\_task} \gets \text{copy(task)}$
    \State $\text{new\_task.start\_time} \gets \text{new\_start\_time}$
    \State $\text{new\_task.end\_time} \gets \text{new\_start\_time} + \text{task.duration}$
    \State \Return $\text{new\_task}$
\EndFunction

\end{algorithmic}
\end{algorithm}

\begin{algorithm}
\caption{Entity Class}\label{alg:entity_class}
\begin{algorithmic}[1]
\State \textbf{Class} \textsc{Entity}
\State \textbf{Private Fields:}
\State $\text{configure} \gets \{\text{hardware performance}\}$
\State $\text{intention} \text{: task}_1\text{, task}_2\text{, ..., task}_n\text{,}$
\State $\text{approved\_intention} \text{: task}_1\text{, task}_2\text{, ..., task}_n\text{,}$
\State $\text{status} \gets \{\text{current status of myself}\}$
\State $\text{connected} \gets \{\text{current sub - system manager and the next sub - system manager}\}$

\Function{Execute}{}
    \While{$\text{True}$}
        \State $t \gets \text{Current\_Time}$
        \State $\text{action} \gets \text{find\_action}(t, \text{approved\_intention})$
        \State $\text{execute}(\text{action})$
    \EndWhile
\EndFunction

\Function{On\_New\_Intention}{}
    \While{$\text{True}$}
        \State $\text{new\_intention} \gets \text{check\_new\_intention}()$
        \State $\text{send}(\text{new\_intention}, \text{connected.sub\_system\_manager})$
    \EndWhile
\EndFunction

\Function{check\_instruction}{}
    \While{$\text{True}$}
        \State $\text{info} \gets \text{check\_info}(\text{connected.sub\_system\_manager})$
        \State $\text{Update}(\text{info}, \text{approved\_intention})$
    \EndWhile
\EndFunction

\Function{RunParallel}{}
    \State $\text{thread1} \gets \text{Thread}(\text{Execute})$
    \State $\text{thread2} \gets \text{Thread}(\text{On\_New\_Intention})$
    \State $\text{thread3} \gets \text{Thread}(\text{check\_instruction})$
    \State $\text{thread1.start()}$
    \State $\text{thread2.start()}$
    \State $\text{thread3.start()}$
\EndFunction
\end{algorithmic}
\end{algorithm}

The intention is then decomposed into tasks, and these tasks need to be shared with the manager well in advance of their execution time. As shown in Fig. \ref{fig:temporal_zone}, the time - task space is divided into five zones: history, frozen, critical, planning, and intention. Only tasks in the intention and planning zones can be submitted to the manager, ensuring that the manager has sufficient time to plan and resolve potential conflicts. This temporal - zone - based mechanism is fundamental for preemptive collaboration in the system.

\subsection{The Sub - System Manager's Role in Conflict Resolution}
Upon receiving task intentions from entities, the sub - system manager's primary responsibility is to resolve conflicts among these tasks. The manager maintains a set of approved tasks for each entity within its spatial range, as defined in the private fields of the \texttt{Sub - System Manager Class} (Algorithm \ref{alg:sub_system_manager}). 

When a new intention arrives, the \texttt{try\_approve} function in the sub - system manager class is invoked. This function iterates through the tasks in the new intention. If a task conflicts with the existing approved tasks, the \texttt{alter} function (Algorithm \ref{alg:alter_function}) is called. The \texttt{alter} function modifies the conflicting task to avoid conflicts, for example, by adjusting its start time to be after the end time of the conflicting task using the \texttt{modify\_task} function within it. 

Once the new intention has been modified to be conflict - free, it is added to the set of approved tasks, and the affected entities are updated. The manager then sends the approved intentions back to the relevant entities.

\subsection{Entity Operations and Collaboration}
Entities in the system have three main functions running in parallel, as defined in the \texttt{Entity Class} (Algorithm \ref{alg:entity_class}). The \texttt{On\_New\_Intention} function continuously checks for new intentions and sends them to the connected sub - system manager as soon as they are detected. The \texttt{check\_instruction} function monitors for updates from the manager and updates the entity's approved intention accordingly. 

The \texttt{Execute} function, on the other hand, is responsible for carrying out the tasks in the approved intention. It retrieves the appropriate action based on the current time and executes it. If an entity moves into a new spatial range, the new manager takes over the approval of its intentions. This well - coordinated interaction between entities and the sub - system manager ensures that the entire system operates efficiently with minimized conflicts.

\section{Application in Road Transportation}
The PHCRTS is designed to offer a holistic global view and
advanced planning by enhancing information sharing across the
transportation network. The system's core lies in the integration
of Road Section Management Units (RSMUs) and Vehicle
Intelligent Units (VIUs)\cite{Tao2024}, which work in tandem to provide
comprehensive traffic management, Simulation experiments in a two-lane
merging scenario demonstrate the effectiveness of PHCRTS,
reducing vehicle time delays by 90\%, increasing traffic capacity by
300\%, and eliminating accidents\cite{10703891}, .

The PHCS achieves its goals by utilizing strategically placed
roadside units that detect and gather data on vehicles and road
conditions, processing and disseminating this information for
improved traffic services. These units also facilitate the
exchange of data messages between vehicles and infrastructure,
leveraging both short-range and long-range communication
technologies.

At the heart of the PHCRTS is a unique processing unit that
expands upon the capabilities of traditional roadside units. This
unit processes real-time data on vehicle status, intentions, and
road infrastructure, and shares this information within its
jurisdiction. This creates a real-time information sharing
network that allows for preemptive cloud uploads of section data.
Vehicles can download this data and establish direct connections
with the processing unit upon entering its area, thus streamlining
the data link establishment process.

Vehicles equipped with specialized units ensure smooth
communication with roadside infrastructure by uploading
vehicle data and receiving information. This setup forms a
seamless, distributed real-time information sharing mechanism
that simplifies data exchange and coordination.

The PHCRTS also incorporates various monitoring devices that
provide road infrastructure information to the processing unit,
which then analyzes and forecasts based on driving intentions,
traffic data, and road conditions. The insights are shared with
vehicles in real time, enhancing traffic management.

To reduce the costs of direct vehicle-to-vehicle
communication, the PHCRTS proposes a shared information
transmission mechanism that relies on the roadside processing
unit for information exchange. This approach has been shown to
significantly reduce information transmission overhead and
delay, particularly when the number of vehicles is large.

By establishing an environmental perception and
information interaction foundation, the PHCRTS provides essential
safety information support for drivers. The system's focus on
global view planning through its unique components enables
efficient coordination and intelligent management of
transportation systems, ensuring a more responsive and safe
traffic environment.

\subsection{Pre-Planned  Spatiotemporal Trajectory}

By employing the PHCRTS, the planning of transportation
vehicle trajectories can be significantly enhanced, leading to
notable improvements in both traffic safety and efficiency. A
key element within the overall architecture is responsible for
gathering an extensive array of real-time data on diverse vehicle
parameters. This includes the vehicle's precise position, current
speed, acceleration patterns, direction of travel, turning angles
during lane changes, and braking status. This data is transmitted
via advanced vehicle-to-infrastructure communication channels
to a central processing entity.

The central processing entity then undertakes a
comprehensive analysis of the aggregated data. Key
considerations in this analysis include the safety distances
between vehicles and the potential for immediate conflicts. By
evaluating dynamic factors such as changes in vehicle speeds,
directions, and traffic patterns, it is able to predict possible
conflicts before they occur. This in-depth analysis forms the
basis for determining the need for pre-programmed trajectories.

In the event that adverse conditions are detected based on the
data analysis, a collaborative control strategy is activated. This
strategy involves the utilization of complex algorithms and
models to calculate the optimal driving trajectory for each
vehicle. Taking into account current traffic conditions, vehicle
characteristics, and potential obstacles, the determined
trajectories are communicated to the relevant vehicle's control
unit. This enables vehicles to be aware of each other's intended
paths, facilitating coordinated driving and reducing the
likelihood of collisions.

For highway on-ramp merge areas, which are typically
hotspots for vehicle conflicts, the system conducts an assessment
to determine whether a ramp vehicle merging into the mainline
without coordinated control will conflict with mainline vehicles.
If no conflict is detected, the ramp vehicle can safely merge
without the need for adjustments. However, if a potential
conflict is anticipated, coordinated control methods, such as
mainline priority and ramp priority control, are implemented. By
pre-programming corresponding vehicle trajectories based on
these methods, ramp and mainline vehicles can adjust their
speeds and positions in a coordinated manner, ensuring a smooth
merge and minimizing the risk of collisions.

When operating in various road conditions, ground-based
radar systems installed on slopes and roadbeds, along with
sensors under bridges, continuously monitor the road surface
and surrounding environment. The monitored data is transmitted
to a central monitoring station, which then disseminates the
information. The system analyzes and processes this data using
advanced algorithms and machine learning techniques to
anticipate potential structural issues such as roadbed subsidence,
slope collapse, or bridge failure. Armed with this knowledge, the
system can pre-program new trajectories to avoid hazards and
maintain driving safety and comfort. For example, if a pothole
or a section of damaged road is detected, the system can redirect
vehicles to alternate routes or adjust their trajectories to
circumvent the hazard. These proactive measures are essential
for maintaining high standards of driving comfort and safety and
for preventing potential threats to vehicular safety.

\begin{algorithm}
\caption{Check New Vehicle}
\begin{algorithmic}[1]
\Function{check\_new\_vehicle}{t\_vehicle, additional\_space}
    \State vehicle\_length, time, speed, x $\gets$ getLength(t\_vehicle), getTime(), getSpeed(t\_vehicle), getDistance(t\_vehicle)
    \State space\_len $\gets$ vehicle\_length + additional\_space
    \If{t\_vehicle[0] == 'm'}
        \State m\_lanes\_info $\gets$ obtain\_lanes\_info(t\_vehicle)
        \State m\_trajectory $\gets$ compose\_trajectory(t\_vehicle, from\_time=time)
        \State m\_list.append({'vehicle': t\_vehicle, 'trajectory': m\_trajectory})
        \If{len(merge\_list) == 0}
            \State merge\_list.append({'vehicle': t\_vehicle, 'trajectory': m\_trajectory})
            \State scheduled\_trajectories[t\_vehicle] $\gets$ complete\_trajectory(m\_trajectory)
        \Else
            \State merge\_into(t\_vehicle, m\_trajectory, additional\_space=additional\_space)
        \EndIf
    \EndIf
    \If{t\_vehicle[0] == 'r'}
        \State r\_lanes\_info $\gets$ obtain\_lanes\_info(t\_vehicle)
        \State r\_trajectory $\gets$ compose\_trajectory(t\_vehicle, from\_time=time)
        \State r\_list.append({'vehicle': t\_vehicle, 'trajectory': r\_trajectory})
        \If{len(self.merge\_list) == 0}
            \State merge\_list.append({'vehicle': t\_vehicle, 'trajectory': m\_trajectory})
            \State scheduled\_trajectories[t\_vehicle] $\gets$ complete\_trajectory(r\_trajectory)
        \Else
            \State merge\_into(t\_vehicle, r\_trajectory, additional\_space=additional\_space)
        \EndIf
    \EndIf
\EndFunction
\end{algorithmic}
\end{algorithm}

\begin{figure}[H]
    \centering
    % First row
    \begin{subfigure}[b]{0.45\textwidth}
        \centering
        \includegraphics[width=\textwidth]{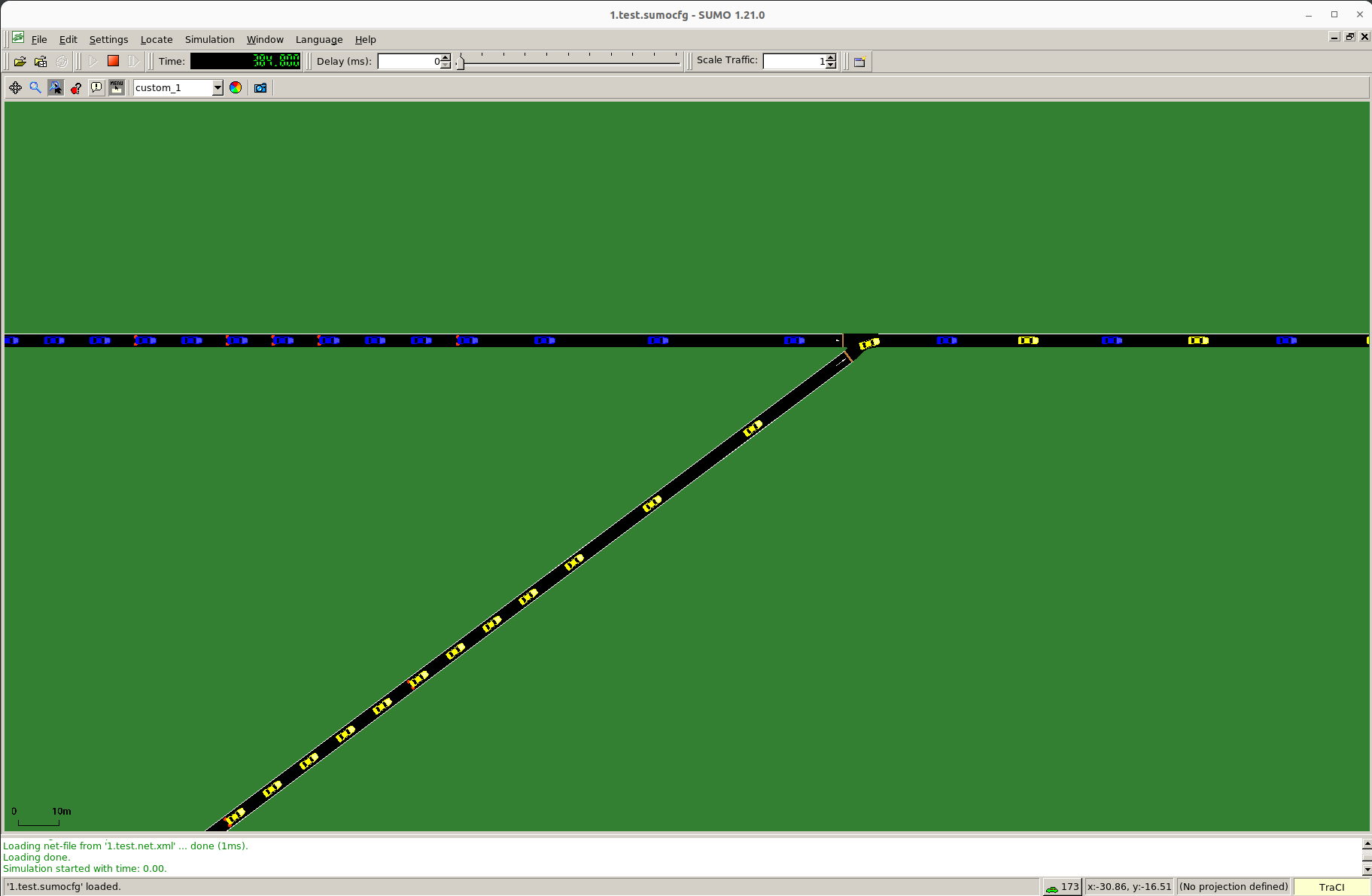}
        \caption{}
        \label{sim:a}
    \end{subfigure}
    \hfill
    \begin{subfigure}[b]{0.45\textwidth}
        \centering
        \includegraphics[width=\textwidth]{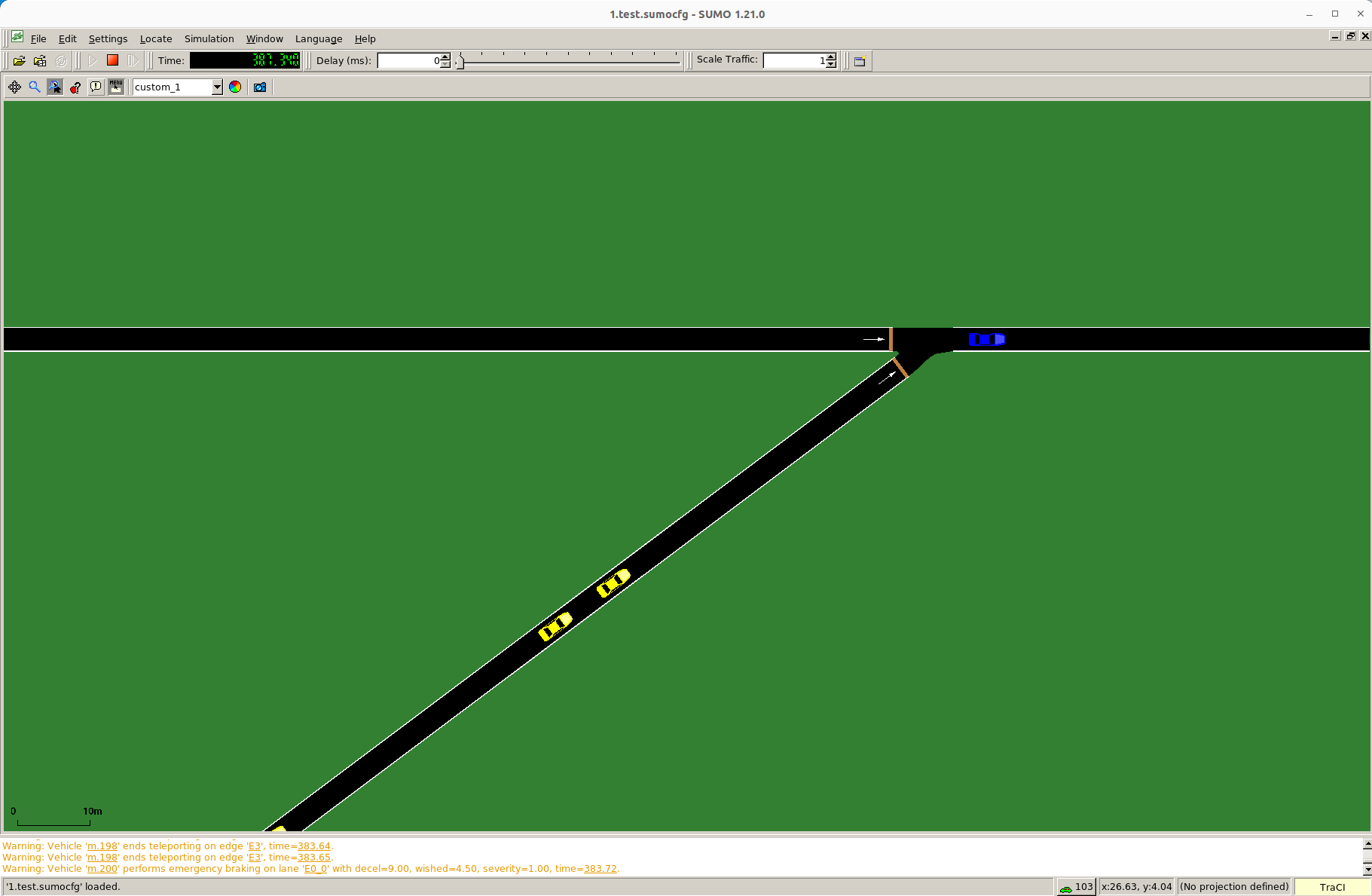}
        \caption{}
        \label{sim:b}
    \end{subfigure}

    % Second row
    \vspace{1em}
    \begin{subfigure}[b]{0.45\textwidth}
        \centering
        \includegraphics[width=\textwidth]{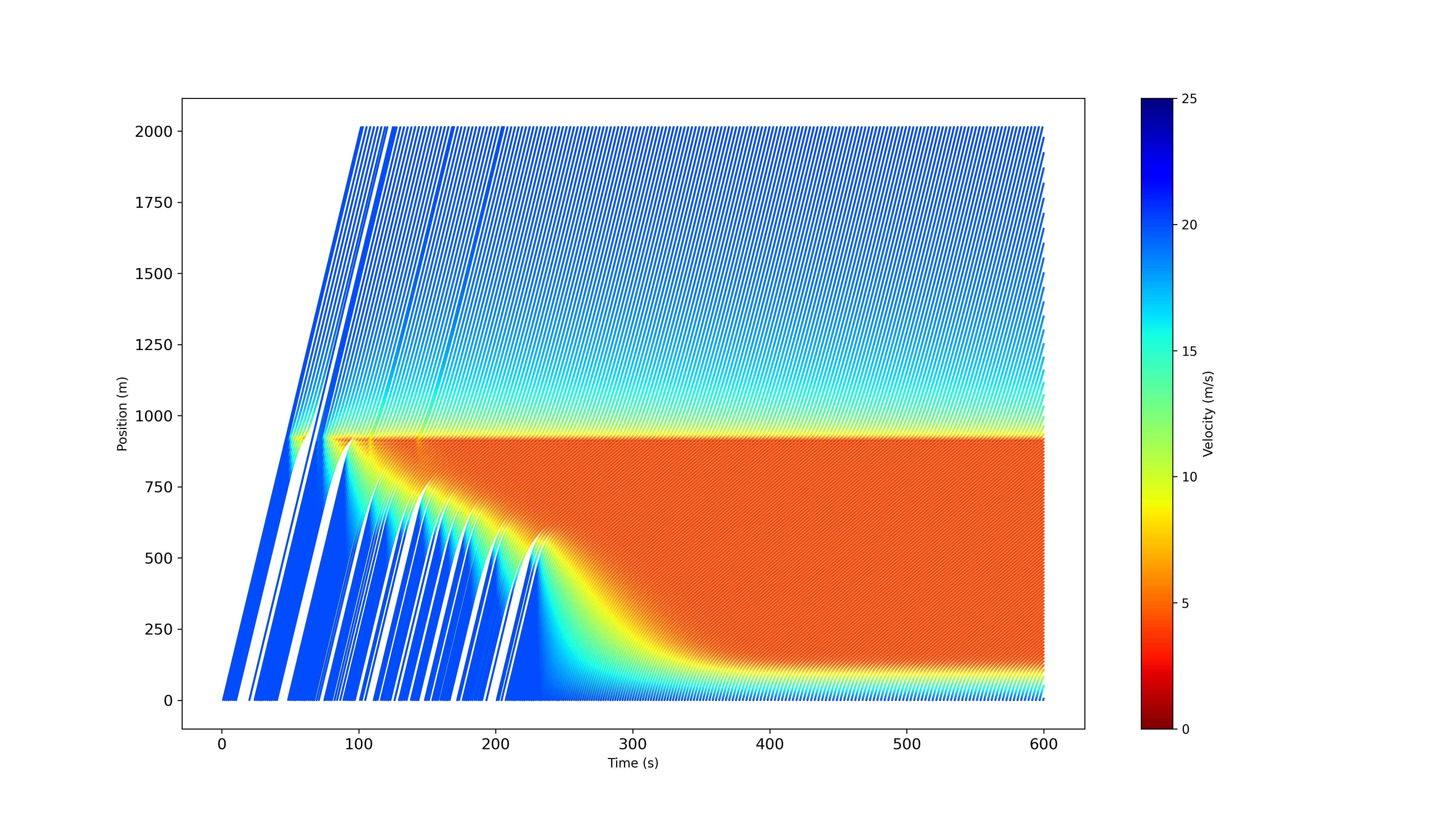}
        \caption{}
        \label{sim:c}
    \end{subfigure}
    \hfill
    \begin{subfigure}[b]{0.45\textwidth}
        \centering
        \includegraphics[width=\textwidth]{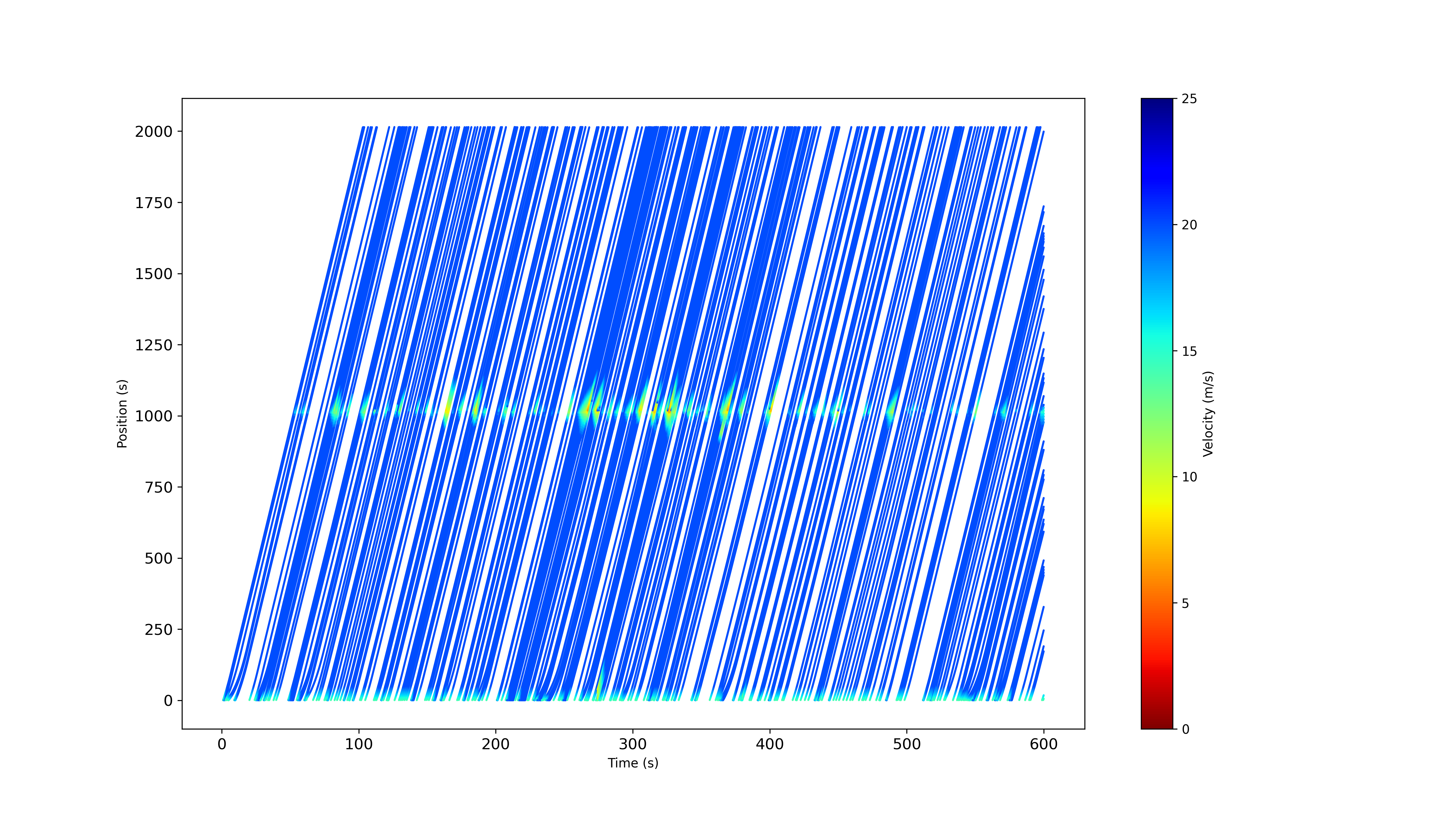}
        \caption{}
        \label{sim:d}
    \end{subfigure}

    % Third row
    \vspace{1em}
    \begin{subfigure}[b]{0.45\textwidth}
        \centering
        \includegraphics[width=\textwidth]{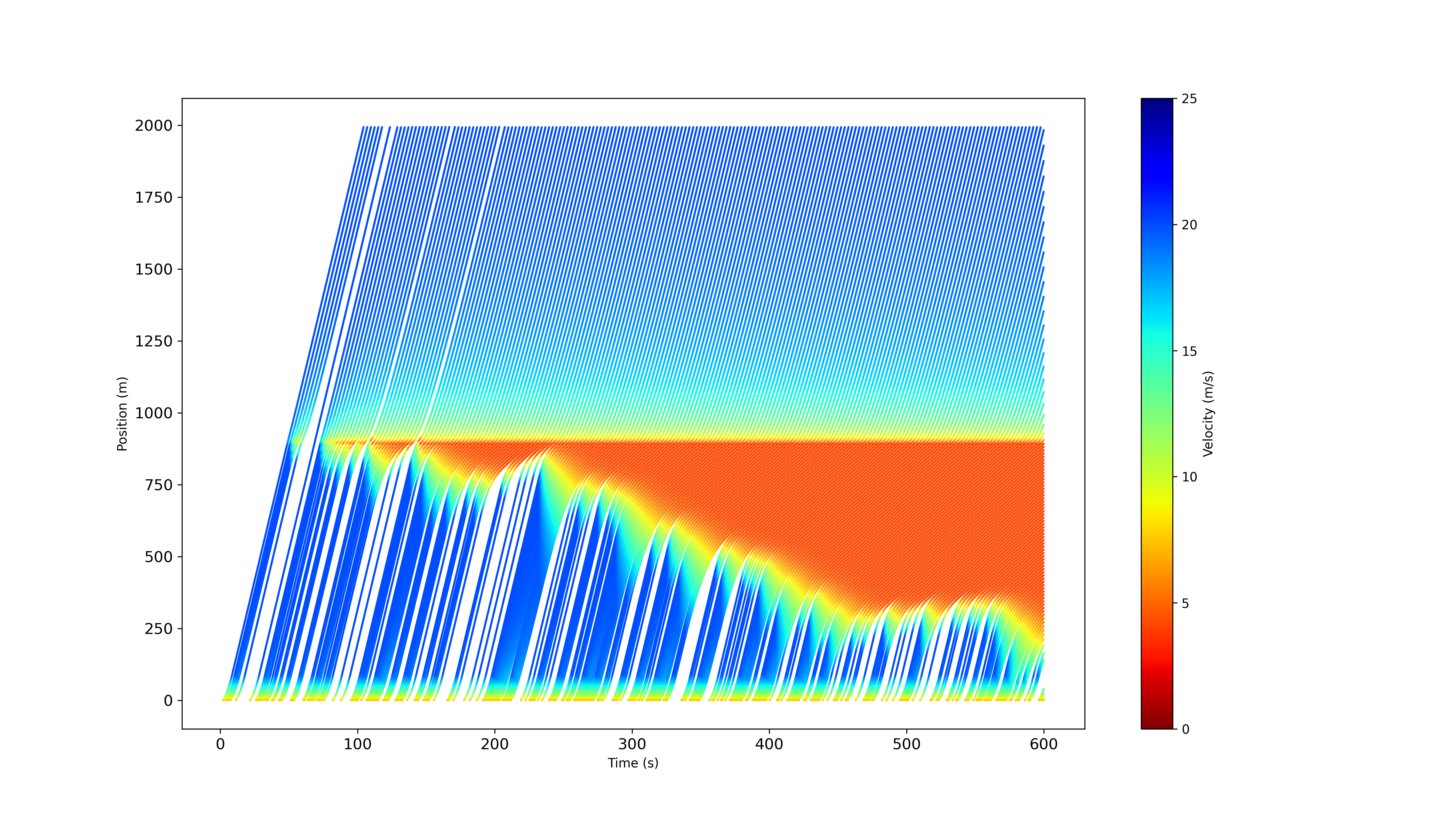}
        \caption{}
        \label{sim:e}
    \end{subfigure}
    \hfill
    \begin{subfigure}[b]{0.45\textwidth}
        \centering
        \includegraphics[width=\textwidth]{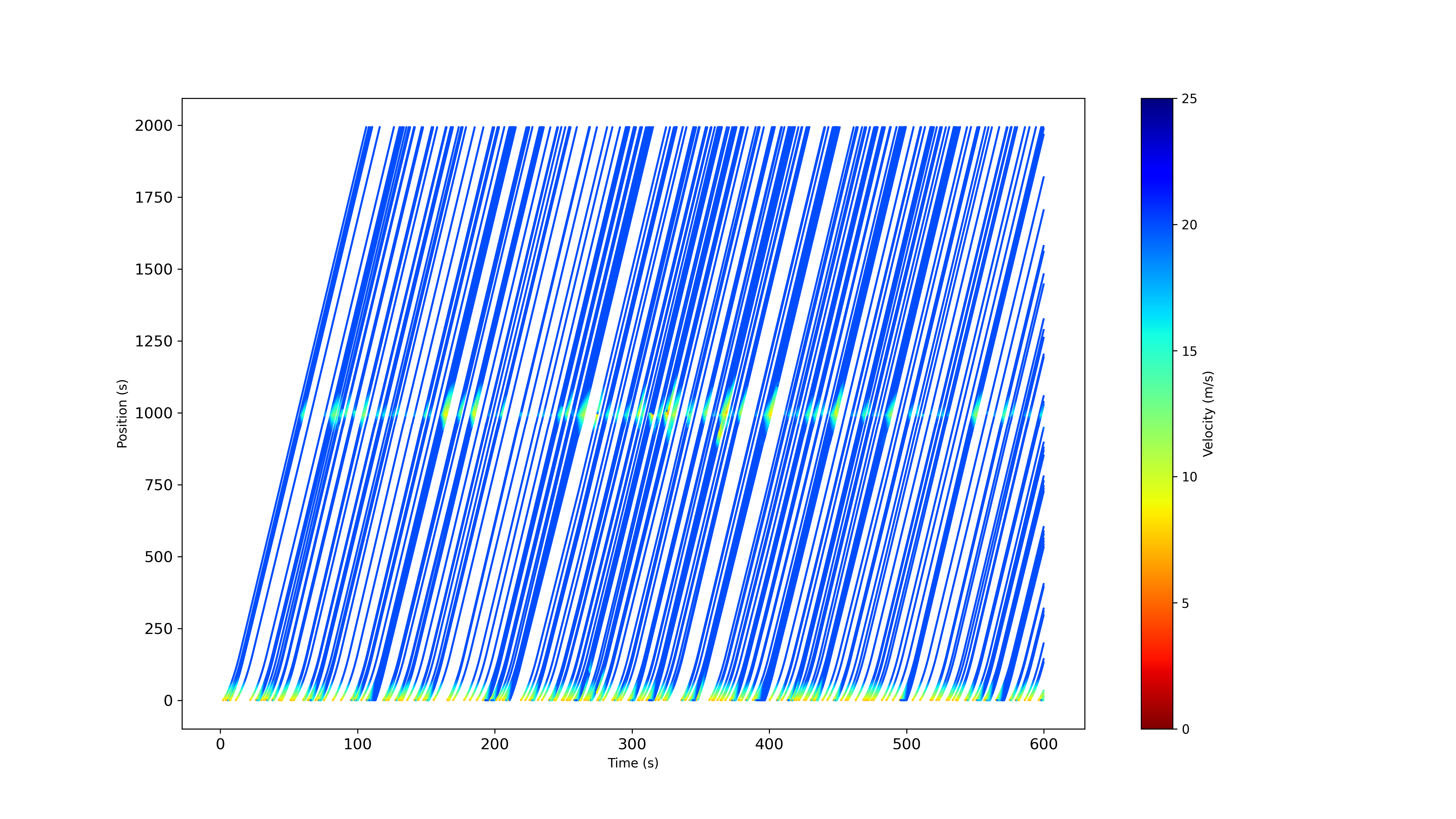}
        \caption{}
        \label{sim:f}
    \end{subfigure}

    \caption{Simulation of Krauss/LC2013 and PHCRTS. \textbf{a}, Krausss/LC2013 Simulation Screenshot. \textbf{b}, Preemptive Collaborative Strategy Simulation Screenshot. \textbf{c}, Mainline Vehicle Operational Status of Krauss/LC2013 Strategy. \textbf{d}, Mainline Vehicle Operational Status of  Preemptive Strategy. \textbf{e}, Ramp Vehicle Operational Status of Krauss/LC2013 Strategy. \textbf{f}, Ramp Vehicle Operational Status of Preemptive Strategy.}
    \label{fig:simulation}
\end{figure}

\subsection{The Effects on Transportation}

The PHCRTS introduces an array of noteworthy effects in the
realm of transportation management. This system attains a
comprehensive perspective through an intricate sharing
mechanism. It collects real-time data on essential vehicle
parameters—including position, speed, acceleration, direction,
turning angles during lane changes, and braking status, and
relays this information to a centralized processing unit.
Consequently, a thorough understanding of the transportation
network is formulated. Complementary to this, ground-based
radar systems, sensors, and vehicle-to-infrastructure
communication further enrich this holistic view by furnishing
insights into road conditions and potential hazards. This
collaborative sharing enables a more precise evaluation of traffic
flow and potential conflicts, empowering PHCRTS to pre-plan
optimal trajectories for vehicles.

The pre-planning of vehicle trajectories has a significant
impact on mitigating traffic delays. By anticipating potential
congestion points and adverse conditions, the PHCRTS can reroute
vehicles or adjust their speeds and paths to circumvent delays.
This proactive strategy minimizes idling time and reduces the
incidence of stop-and-go traffic, leading to a smoother traffic
flow and notably lower delays.

Furthermore, the PHCRTS substantially enhances traffic
capacity. Leveraging its comprehensive understanding of the
transportation network, it optimizes road space utilization by
pre-planning trajectories that maximize vehicle movement
efficiency. For instance, in highway on-ramp merge zones, the
PHCS implements coordinated control methods ensuring
seamless integration of ramp vehicles into mainline traffic
without disrupting flow or reducing capacity. By efficiently
distributing vehicles across the road network, PHCRTS
accommodates a greater volume of traffic, thereby bolstering
capacity.

In addition to addressing delays and improving capacity, it
also elevates traffic safety. The pre-planned trajectories facilitate
mutual awareness among vehicles regarding intended paths,
reducing the risk of collisions. This is particularly vital in areas
with intricate traffic interactions, such as on-ramp merges.
Additionally, the PHCS's proactive detection and response to
road conditions help prevent accidents caused by hazards like
potholes, road defects, and structural issues. By pre-
programming alternative trajectories to bypass these hazards,
PHCS ensures the safety of drivers and passengers.

\subsection{Strategy of Applications in Road Transportion System}

The PHCRTS represents a significant leap forward in
optimizing traffic flow and enhancing safety enabling real-time
driving intention sharing and pre-planned trajectory
coordination among vehicles. This system not only addresses
traditional bottlenecks such as merging and lane-changing
scenarios but also extends its preemptive collaborative
capabilities to a wide range of transportation applications. In this
paper, we explore several innovative applications of PHCS in
transportation systems, emphasizing its potential to
revolutionize urban mobility. In this way, PHCRTS is formed.

\subsubsection{Merging and Lane-Changing Scenarios}

The PHCRTS revolutionizes merging and lane-changing
scenarios by leveraging real-time data sharing and advanced
predictive analytics. By constructing an overall view of traffic
conditions, the PHCS can anticipate merging bottlenecks and
lane-changing conflicts, enabling proactive management of
these critical situations. The system integrates data from various
sensors, cameras, and GPS devices, allowing it to predict driver
intentions and vehicle movements with high accuracy. This
predictive capability enables the PHCRTS to suggest optimal
merging and lane-changing strategies, reducing congestion,
enhancing traffic flow, and minimizing the risk of accidents.
Furthermore, PHCRTS can communicate these strategies to drivers
through in-vehicle displays or smart traffic signs, facilitating
smoother and safer transitions between lanes and merging points.

\subsubsection{Improved Safety and Reduced Accidents}
Safety is a paramount concern in transportation systems, and
the PHCRTS contributes significantly to enhancing it. By analyzing
historical accident data and real-time traffic conditions, PHCS
can identify high-risk areas and times, allowing authorities to
deploy additional resources such as police patrols or safety
barriers. Furthermore, the system can send alerts to drivers about
potential hazards, such as icy roads or construction zones,
helping them make informed decisions and avoid accidents. The
integration of connected vehicle technology, where vehicles
communicate with each other and infrastructure, further
enhances safety by enabling preemptive braking, lane-keeping
assistance, and other collision avoidance measures.

\subsubsection{Traffic Signal Optimization}
By leveraging real-time data and predictive algorithms, the
PHCS significantly enhances traffic signal optimization through
dynamic timing adjustments. By gaining a comprehensive view
of traffic conditions across the network, the system identifies
patterns and trends, allowing it to predict future traffic flows
with high accuracy. This predictive capability enables the PHCS
to adjust signal timings in real-time, optimizing green and red
light durations to meet evolving traffic demands. Consequently,
the traffic signal system becomes more efficient and responsive,
reducing delays, improving road capacity utilization, and
enhancing overall traffic flow. Furthermore, the PHCS can
integrate data from connected vehicles, refining signal timings
even further to accommodate real-time traffic conditions and
driver behaviors.

\subsubsection{Emergency Vehicle Priority}
Leveraging real-time data sharing and predictive analytics,
the implementation of a Priority Traffic Control System (PHCRTS)
is vital for facilitating emergency vehicle priority. By
constructing a comprehensive view of traffic conditions, this
system identifies emergency vehicles and prioritizes their routes,
guaranteeing swift arrival at their destinations. Dynamically
adjusting traffic signals, the system grants green lights to
emergency vehicles as they approach intersections, minimizing
delays and improving response times. Additionally, it
communicates real-time updates to other drivers, alerting them
to the presence of emergency vehicles and encouraging them to
yield. This proactive method ensures efficient navigation for
emergency services through traffic, ultimately saving lives and
reducing the impact of emergencies.

\subsubsection{Traffic Incident Management}
The implementation of the PHCRTS fundamentally transforms
traffic incident management by utilizing real-time data sharing
and predictive analytics to provide swift and effective responses
to traffic disruptions. By gaining a comprehensive view of traffic
conditions, this system detects incidents such as accidents, road
closures, or hazardous conditions with remarkable speed and
precision. It then communicates real-time updates to drivers,
suggesting alternative routes and steering them away from
congested areas. Furthermore, the system prioritizes emergency
vehicle routes, ensuring rapid arrival at the incident site.
Integration with IoT devices, including sensors and drones,
allows for real-time damage assessments and accelerated
recovery efforts. This proactive and holistic approach ensures
efficient incident management, minimizing delays and
enhancing overall road safety.

\subsubsection{Revolutionizing Logistics and Supply Chain}

In the field of logistics and supply chain management,
multiple independent entities such as manufacturers, distributors,
and transportation companies operate. PHCS can facilitate
information sharing among them, enabling accurate prediction
of each entity's actions and preemptive planning of the flow of
goods. This leads to significant improvements in efficiency,
reducing delivery times and minimizing inventory costs. For
example, manufacturers can share production plans with
distributors in advance, and transportation companies can plantheir routes and schedules accordingly, ensuring a seamless flow
of products from production to consumption.

\section{Conclusions}

Within the road traffic system, the implementation of the
preemptive collaborative system can enhance multi-vehicle
coordination by acquiring operational information and driving
intentions of vehicles within perceived areas (such as weaving
zones, entry ramps, and exit ramps). This enables the planning
of vehicle trajectories and the transmission of control commands
to the vehicles. Furthermore, in the absence of traffic signals, it
ensures efficient vehicle passage by developing collaborative
passage strategies based on intersection lane layouts, real-time
traffic data, and vehicle intentions, thereby ensuring the safety
and efficiency of different directional vehicle queues at
unsignalized intersections. Regarding comprehensive guidance
for emergency rescue, continuous monitoring of road
operational information can proactively identify accident types
and, utilizing vehicle information, road data, and information
from third-party rescue units, generate optimal routes and traffic
strategies for rescue vehicles, guiding other vehicles to optimize
their paths to facilitate successful rescue operations.

The application in maritime traffic can enhance navigation
safety and efficiency. In port management, preemptive path
planning aids in scheduling vessels’ entry and exit, improving
terminal operation efficiency and reducing waiting times.
Through vehicle-road collaboration technologies, real-time
information is shared among vessels and with relevant entities
such as ports and the Coast Guard, enhancing collaborative
operational capabilities and improving response to emergencies.

In urban traffic management, the use of preemptive
collaborative technologies can optimize traffic signal control,
reduce congestion, and improve passage efficiency. By
analyzing passenger flow data, public transport routes and
station setups can be optimized, enhancing the convenience and
attractiveness of public transit and encouraging citizens to utilize
it. In the event of emergencies (such as traffic accidents or
natural disasters), rapid adjustments to traffic flow and
emergency response routes can significantly enhance rescue
efficiency.

\section*{Declarations}
\begin{itemize}
\item  Competing interests: The authors declare that they have no competing interests. None of the researchers have any financial, personal, or professional relationships that could potentially influence or bias the research, its interpretation, or the reporting of results.
\item Code availability: Related code of this research is available at \href{https://github.com/tpeng1977/PreemptiveHolisticCollaborativeSystem}{github}.
\end{itemize}


\bibliography{bibliography}% common bib file

\end{document}